# SHARC: SHAP-Based Interpretability in Machine Learning Risk Models for Regulatory Capital under ICAAP and CCAR

**Ujjwala Vadrevu**

Replication code: https://colab.research.google.com/drive/1nrlSqmG10DNerNmEqGIh3EB9CcLWIgH9

## Abstract

The adoption of non-parametric machine learning models in regulatory capital estimation introduces a fundamental governance challenge: the inability to explain model outputs in a manner auditable by supervisory bodies. This 'black box' problem is a primary barrier to the adoption of Gaussian Process Regression (GPR) and related ML architectures in Internal Capital Adequacy Assessment Process (ICAAP) and Comprehensive Capital Analysis and Review (CCAR) workflows, despite their demonstrated predictive superiority over traditional parametric approaches.

This paper resolves this barrier by introducing SHARC (SHAP for Regulatory Capital), a structured explainability framework for the Hybrid GPR-HS architecture building on earlier work and its stress-testing extension. SHapley Additive exPlanations (SHAP), derived from cooperative game theory and satisfying the axiomatic properties of Local Accuracy, Missingness, Consistency, and Efficiency, are applied to Stressed Value-at-Risk (SVaR) outputs under three macro scenarios: West Asia War, Climate Risk, and AI Bubble/Regulatory Burden.

SHARC formalizes the decomposition of SVaR into baseline, mean-driven, and volatility-driven components, enabling a transparent mapping between scenario design and resulting capital levels. The analysis produces two principal findings. First, the SHARC decomposition consistently links non-linear SVaR outputs to the underlying scenario inputs, confirming framework fidelity and enabling auditable traceability of capital drivers. Second, under stress conditions, the mean return component (directional loss magnitude) dominates the variance component (volatility baseline) in determining capital levels, revealing a non-linear structure with direct implications for capital limit-setting, position management, and hedging strategy design. The results establish SHARC as a regulator-aligned explainability layer that transforms the Hybrid GPR-HS framework into a fully auditable capital engine, consistent with the transparency requirements of FRTB, ICAAP Pillar 2, and CCAR model documentation standards.

## 1. Introduction

### 1.1 The Black Box Problem in Regulatory Finance

The progressive adoption of machine learning methods in financial risk modeling has created a fundamental tension with the governance requirements of regulatory capital frameworks. Regulators, including the Basel Committee, the Federal Reserve, and the European Central Bank; require not only that models produce accurate risk estimates, but that those estimates be explainable: decomposable into underlying risk drivers in a manner that a non-technical supervisor can interrogate, challenge, and approve (*Dodd-Frank Wall Street Reform and Consumer Protection Act*, 2010; Basel Committee on Banking Supervision [BCBS], 2019)

Traditional parametric models like GARCH-family and Historical Simulation approaches, satisfy this requirement by construction. Their functional forms are fully specified, and the contribution of each input to the output can be derived analytically. Non-parametric machine learning models, including Gaussian Process Regression, lack this property. The GPR posterior predictive distribution is an integral over all admissible functions weighted by the kernel-defined prior, producing outputs whose attribution to specific inputs is non-trivial and non-linear (Rasmussen and Williams, 2005). This renders GPR a 'black box' from the perspective of regulatory auditability, despite its demonstrated predictive superiority.

The Fundamental Review of the Trading Book (FRTB) explicitly requires model transparency and documentation sufficient for regulatory review (Basel Committee on Banking Supervision [BCBS], 2019). ICAAP Pillar 2 guidance mandates that institutions demonstrate the 'logic' behind their capital estimates to supervisory bodies. CCAR model documentation standards require auditable attribution of capital projections to specific scenario inputs. These requirements collectively create a regulatory barrier to GPR adoption that is independent of the model's predictive performance.

## 1.2 SHAP as the Regulatory Solution

SHapley Additive exPlanations (SHAP), developed by Lundberg and Lee (2017) from the game-theoretic Shapley value framework (Shapley, 1953), provides the only model-agnostic explainability methodology that satisfies all four axiomatic requirements for fair, consistent attribution: Local Accuracy, Missingness, Consistency, and Efficiency. These axioms are not merely technical, they correspond directly to the properties that regulatory audit processes require of model documentation: completeness, absence of phantom attribution, monotonicity, and zero-remainder decomposition.

This paper develops SHARC (SHAP for Regulatory Capital), a structured application of SHAP to the Hybrid GPR-HS framework established in Vadrevu, (2026a) and extended to forward-looking stress scenarios in Vadrevu, (2026b). SHARC formalizes the decomposition of Stressed Value-at-Risk (SVaR) into baseline, mean-driven, and volatility-driven components, enabling a transparent mapping between scenario design and the resulting capital requirement.

The SHAP framework is applied to SVaR outputs generated under three macro scenarios: West Asia War, Climate Risk, and AI Bubble/Regulatory Burden. The choice to focus on SVaR rather than base case VaR is deliberate: SVaR represents the capital-binding regulatory measure, and explainability is most valuable precisely where capital requirements are highest. The SHARC formulation therefore provides a regulator-aligned interpretation of capital structure, linking model outputs directly to their underlying drivers in a manner suitable for supervisory review.

## 1.3 Contributions

This paper makes four primary contributions:

1. Axiomatic Framework for Regulatory Explainability: Formally maps the four SHAP axioms to their regulatory counterparts in FRTB, ICAAP, and CCAR documentation requirements, establishing SHAP as a theoretically grounded foundation for model explainability in regulatory capital frameworks.
2. SHARC Framework for Regulatory Capital Decomposition: Introduces SHARC (SHAP for Regulatory Capital), a structured decomposition of SVaR into baseline, mean-driven, and volatility-driven components, enabling a transparent and regulator-aligned interpretation of capital drivers within the Hybrid GPR-HS architecture.
3. Empirical Validation Across Stress Scenarios: Demonstrates that the SHARC decomposition consistently aligns SVaR outputs with the underlying scenario design across three macro scenarios, confirming framework fidelity and enabling auditable traceability of capital drivers.

4. Practical Risk Governance Template: Translates the SHARC-based decomposition into actionable recommendations for capital limit-setting, position management, model monitoring, and regulatory reporting, operationalizing explainability for production risk management workflows.

# 2. Literature Review

## 2.1 The Regulatory Demand for Model Transparency

The regulatory imperative for model transparency in financial risk management predates the machine learning era but has intensified with the proliferation of complex, non-linear modelling approaches. The Dodd-Frank Wall Street Reform and Consumer Protection Act (2010) established the legal foundation for model risk governance in the United States, requiring that models used in capital determination be documented, validated, and explainable. SR 11-7, the Federal Reserve's model risk management guidance, defines a 'model' as any quantitative method whose output informs business decisions and requires that model logic be documented, tested, and understood by both users and validators (Federal Reserve, 2020).

Under the Fundamental Review of the Trading Book (FRTB), the internal model approach (IMA) requires institutions to demonstrate that risk models are sufficiently transparent for supervisory benchmarking and internal audit (Basel Committee on Banking Supervision [BCBS], 2019). ICAAP Pillar 2 guidance from the European Banking Authority (EBA) and the Prudential Regulation Authority (PRA) similarly requires that institutions explain the drivers of their capital requirements in terms accessible to non-technical senior management and supervisory reviewers.

Collectively, these requirements define a functional specification for explainability in regulatory capital models: the ability to decompose capital outputs into underlying risk drivers in a manner that is complete, non-phantom, monotonic, and free of residual components.

## 2.2 The Evolving XAI Landscape

The field of Explainable Artificial Intelligence (XAI) has produced several approaches to the explanation problem, including LIME (Local Interpretable Model-agnostic Explanations), integrated gradients, attention mechanisms, and SHAP (Barredo Arrieta *et al.*, 2020). Among these, SHAP is particularly well suited for regulatory applications due to its axiomatic grounding. Lundberg and Lee (2017) demonstrated that SHAP is the only additive feature-based explanation method satisfying the properties derived from the Shapley value framework (Shapley, 1953): dummy player, symmetry, additivity and efficiency. These properties correspond closely to the requirements of regulatory model governance, including completeness, consistency, and zero-remainder decomposition.

Guidotti et al. (2018) provide a comprehensive taxonomy of XAI methods, noting that local explanation approaches - those that interpret individual predictions rather than aggregate model behaviour - are most relevant for high-stakes regulatory applications where specific capital calculations must be justified. This distinction is directly applicable in the ICAAP and CCAR context, where regulatory review focuses on point-in-time capital outputs rather than statistical summaries of average model behaviour.

Gramegna and Giudici (2021) demonstrate the application of SHAP in financial risk modelling, highlighting its usefulness for regulatory reporting. However, their analysis is confined to credit risk models with relatively simple feature structures. This paper extends the application to a non-parametric Gaussian Process Regression (GPR) framework, where the posterior predictive structure does not admit analytical gradient-based explanation. In this setting, the model-agnostic Kernel SHAP approach is required to enable a consistent decomposition of model outputs.

## 2.3 The Shapley Value Framework

The theoretical foundation of SHAP lies in cooperative game theory. The Shapley value, introduced by Shapley (1953), provides the unique solution to the problem of fairly distributing a coalition's total payoff among its members, satisfying four axioms: efficiency (the total payoff is fully distributed), symmetry (equivalent players receive equal shares), dummy player (non-contributing players receive nothing), and linearity (additive games produce additive distributions).

Lundberg and Lee (2017) reformulate the Shapley value as a feature-based explanation method for machine learning predictions. In this context, the 'coalition' is the set of input features, the 'payoff' is the model prediction, and the 'fair distribution' represents the contribution of each feature to the prediction. The resulting SHAP value $\varphi_i$ for feature *i* represents the average marginal contribution of that feature across all possible feature orderings, ensuring that the decomposition is both complete and consistent.

For the GPR-HS framework, the SHAP decomposition is applied to the SVaR calculation function $f(\mu, \sigma) = SVaR$, where the $2N$ input features consist of the *N* GPR-predicted conditional means and *N* predicted conditional standard deviations for all portfolio assets under the stress scenario. The SHARC framework (SHAP for Regulatory Capital) formalizes this decomposition for regulatory capital as:

$$SVaR_{SHARC} = \varphi_0 + \Sigma_i(\varphi_{\mu_i} + \varphi_{\sigma_i})$$

where:

- $SVaR_{SHARC}$ represents the final regulatory capital measure under the SHARC decomposition,
- $\varphi_0$ represents the baseline capital level, which in the stress testing context corresponds to the pre-stress (Day 1) SVaR,
- $\varphi_{\mu_i}$ captures the contribution of the conditional mean (directional loss component),
- $\varphi_{\sigma_i}$ captures the contribution of the conditional variance (volatility component) for index $i$.

This formulation provides a structured decomposition of regulatory capital into baseline, directional, and volatility-driven components, enabling a transparent interpretation of the drivers of SVaR under stress.

## 2.4 Mean vs. Variance in Tail Risk Decomposition

The theoretical relationship between the mean return and the variance in determining tail risk measures has been studied extensively in the risk management literature. McNeil, Frey and Embrechts (2015) establish that under heavy-tailed distributions, extreme quantile measures (VaR and ES) are primarily determined by the distributional tail shape and the scale of the loss, with the mean playing a secondary role under normal conditions. However, under conditioning on extreme negative return paths, as in the SVaR context, the directional magnitude of the loss can become the binding constraint on the capital measure.

This theoretical perspective is consistent with the empirical behaviour observed in this paper. Within the SHARC framework, which decomposes SVaR into mean-driven and volatility-driven components, the mean return component (directional loss) becomes the dominant driver of the final capital output under strong stress conditioning, while the variance component provides a secondary, stabilizing effect.

This decomposition provides a quantitative mechanism for the well-known practitioner observation that "volatility sets the scene, but the loss event pulls the trigger" in tail risk capital calculations.

# 3. Methodology

## 3.1 Framework Foundation

The SHAP analysis is applied to the Hybrid GPR-HS framework documented in Vadrevu, (2026a) and its stress testing extension developed in Vadrevu, (2026b). These works describe the full GPR architecture, kernel specification, Aggressive Noise Initialization (ANI) strategy, scenario construction methodology, Scenario-Averaged Covariance Stabilization (SACS) framework, and SVaR calculation.

This paper focuses on the SHAP-specific methodology, formalised as the SHARC (SHAP for Regulatory Capital) framework, and its interpretation within a regulatory capital context.

## 3.2 SHAP Axiomatic Framework and Regulatory Mapping

The four SHAP axioms are formally mapped to their regulatory counterparts below. This mapping establishes the theoretical case for SHAP as an appropriate explainability tool for regulatory capital model documentation.

**Table 1: SHAP Axiomatic Properties and Regulatory Interpretations**

| Axiomatic Property | Formal Definition | Regulatory Interpretation |
|---|---|---|
| Local Accuracy | The sum of SHAP values equals the model output for the specific prediction instance. | The decomposition is complete: every basis point of SVaR is explained by underlying input components |
| Missingness | Features with zero input value receive a SHAP value of zero. | Features absent from the scenario contribute no component to the capital measure, preventing phantom contributions |
| Consistency | If a model changes so that a feature's marginal contribution increases, its SHAP value cannot decrease. | Ensures monotonicity: a higher-shock input cannot produce a lower contribution to capital. |
| Efficiency | SHAP values sum exactly to the difference between the model output and the baseline value. | The full capital decomposition is distributed across inputs with zero residual, i.e., no unexplained capital remains |

The efficiency property is of particular regulatory significance: it guarantees that the SHAP decomposition produces a complete representation with zero residual. This ensures that every basis point of the terminal SVaR output can be connected to an underlying input component, supporting the ICAAP requirement for full transparency of capital drivers within the SHARC framework.

## 3.3 Input Feature Specification

The SHAP analysis is applied to the SVaR calculation function, treating it as the target function $f$ to be explained. The input features for the analysis consist of the $2N$ outputs from the $N$ stressed GPR models: $N$ predicted conditional means and $N$ predicted conditional standard deviations, one per portfolio index, evaluated at the terminal day T = 252 of each stress scenario path.

**Table 2: SHAP Input Feature Specification**

| Feature Type | Feature Name | Definition | Regulatory Significance |
|---|---|---|---|
| Mean Component | ^GSPC_Mean, ^STOXX50E_Mean, ... (N features) | GPR-predicted conditional mean return for each index under the stressed path | Captures the directional loss magnitude: the ‘event trigger’ during tail events |
| Variance Component | ^GSPC_Sigma, ^STOXX50E_Sigma, ... (N features) | GPR-predicted conditional standard deviation for each index under the stressed path | Captures the volatility contribution: the ‘risk baseline’ setter |

This feature structure separates two conceptually distinct risk channels: the directional loss contribution (mean components) and the volatility contribution (variance components). This separation enables the behaviour of regulatory capital under stress to be decomposed explicitly within the SHARC framework, rather than inferred implicitly.

### 3.4 Kernel SHAP Application

Kernel SHAP introduced by Lundberg and Lee (2017) is the appropriate SHAP variant for this application. Unlike Tree SHAP or gradient-based SHAP methods, which exploit the internal structure of specific model classes, Kernel SHAP is model-agnostic, treating the model as a black box and estimating SHAP values through a weighted linear regression over perturbed input coalitions. This makes it suitable for the GPR architecture, whose posterior predictive function does not admit analytical gradient-based decomposition.

The SVaR calculation function $f(\mu_1, \dots, \mu_N, \sigma_1, \dots, \sigma_N)$ is treated as the target function to be explained. For each stress scenario, the SHAP analysis is performed on the terminal day (T = 252) SVaR output, decomposing the difference between the predicted SVaR and the baseline component $\varphi_0$ into the *2N* feature contributions within the SHARC framework.

The SHAP analysis produces a decomposition of the terminal SVaR output into feature-level components, capturing the contribution of both directional (mean) and volatility (variance) inputs within the SHARC framework. These decompositions form the basis for the analysis of capital drivers under stress scenarios.

### 3.5 Local vs. Global Interpretation

This study deliberately employs local SHAP interpretation, explaining specific, point-in-time SVaR outputs during peak stress, rather than global SHAP which summarizes average model behaviour across historical conditions. This choice is grounded in the regulatory context: ICAAP and CCAR auditors require auditable explanations for individual capital calculations, rather than statistical summaries of average model behaviour. The local interpretation provides granular input-output traceability consistent with regulatory documentation requirements.

Global interpretation - analysing average SHAP values across the full backtesting window - represents a natural extension for future research and could inform model monitoring frameworks and MLOps dashboards. Lundberg et al. (2020) demonstrate how local SHAP values can be aggregated into global importance measures while preserving their axiomatic properties, providing a methodological bridge between the local approach adopted here and the broader governance view required at the portfolio level.

## 4. Results

### 4.1 SHARC Decomposition Maps to Scenario Design

The primary validation of the SHARC framework is the confirmation that the decomposition correctly reflects the scenario's input design: indices receiving the highest cumulative shocks under each scenario are expected to produce the largest contributions to SVaR, while indices receiving more moderate shocks act as offsetting components.

This fidelity check is satisfied consistently across all three scenarios. Under the West Asia War scenario, where European and Hong Kong markets receive the deepest shocks (EURO STOXX 50: -40%; Hang Seng: -45%), the EURO STOXX 50 mean component emerges as the primary driver of the SVaR decomposition. Conversely, the FTSE 100 and Nikkei 225 receiving relatively moderate shocks at -30%, act as mitigating components, with their smaller losses partially offsetting the deeper European and Asian losses.

Under the AI Bubble/Regulatory Burden scenario, where the S&P 500 receives the deepest shock at -45% reflecting high technology sector concentration, the S&P 500 mean component dominates the decomposition. Under the Climate Risk scenario, where geographic shock concentration produces the most heterogeneous cross-market impact, the decomposition pattern reflects the differential regional exposure, with East Asian and European indices contributing the largest positive components to the loss magnitude.

This consistent behaviour across all three scenarios provides strong empirical evidence that the Hybrid GPR-HS framework is not a black box in the regulatory sense: its outputs faithfully transmit the scenario input design through the non-linear GPR architecture, and the SHARC decomposition renders that transmission transparent and interpretable.

## 4.2 Cross-Scenario SHARC Summary

Figures 1 - 3 present the SHARC Summary Plot decomposition for the West Asia War, Climate Risk, and AI Bubble/Regulation scenarios respectively. The plots illustrate the relative contribution magnitudes of the mean and variance components across portfolio indices under each scenario.

**Figure 1: SHARC Summary Plot - West Asia War**

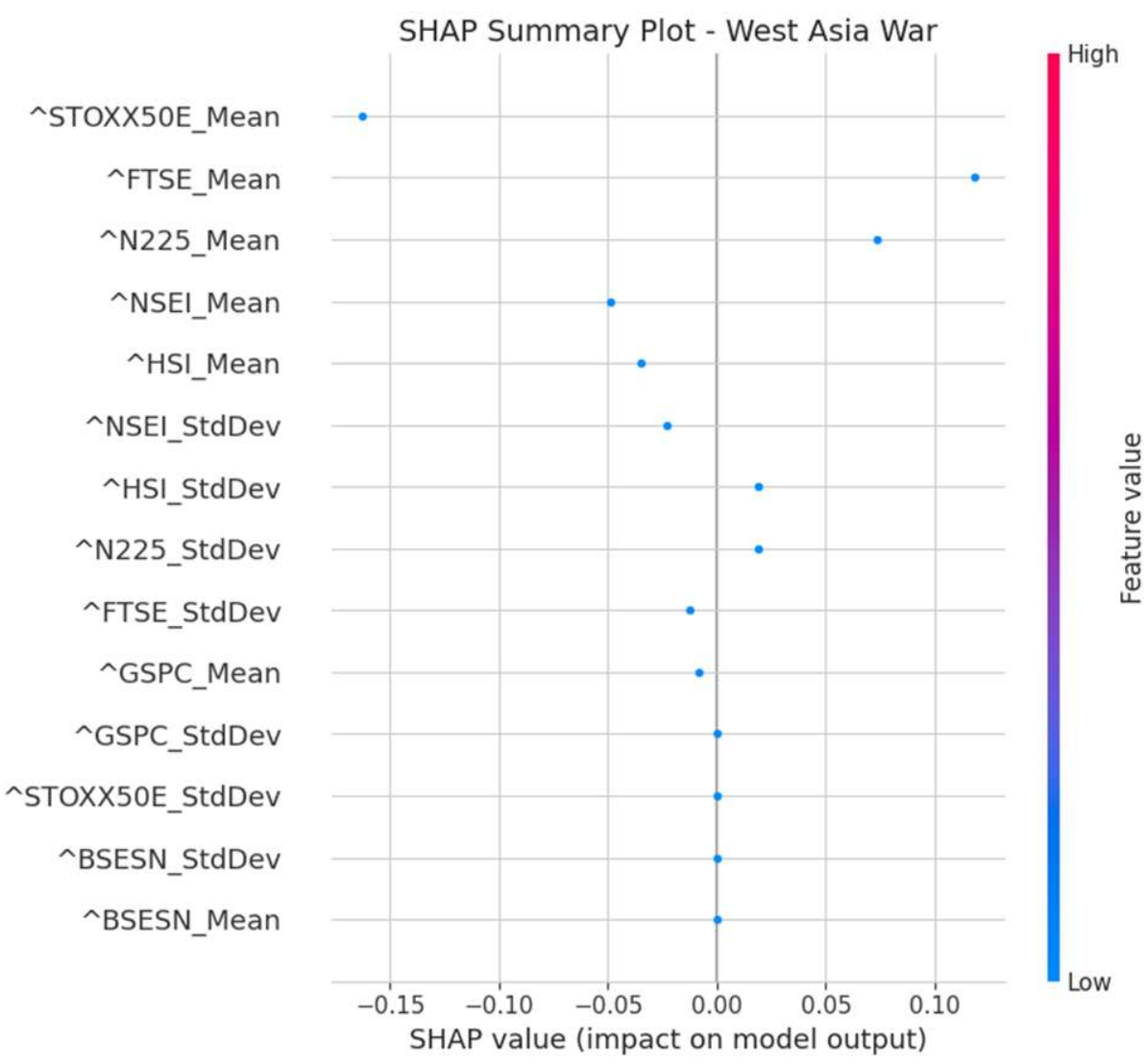

**Figure 2: SHARC Summary Plot - Climate Risk**

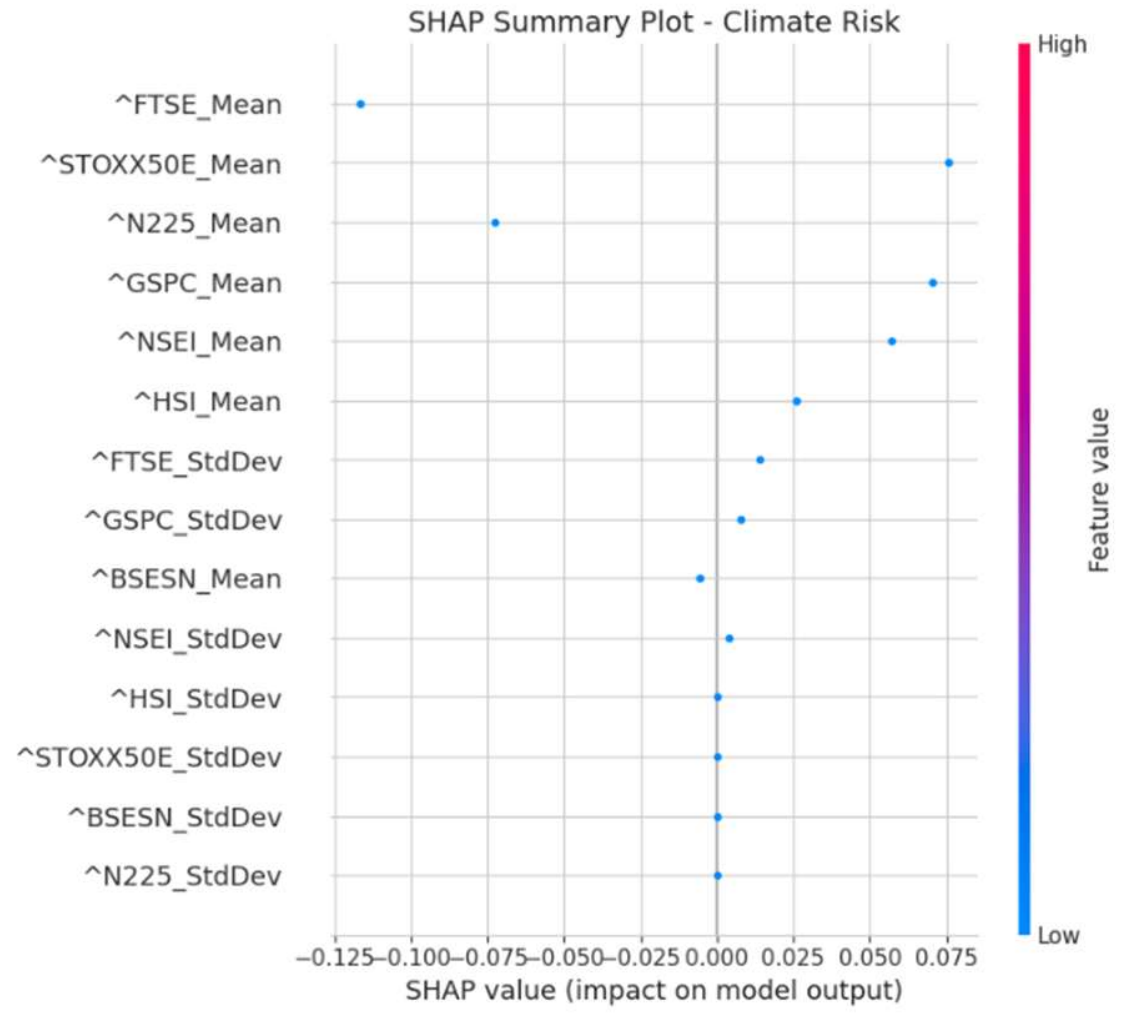

**Figure 3: SHARC Summary Plot – AI Bubble/Regulation**

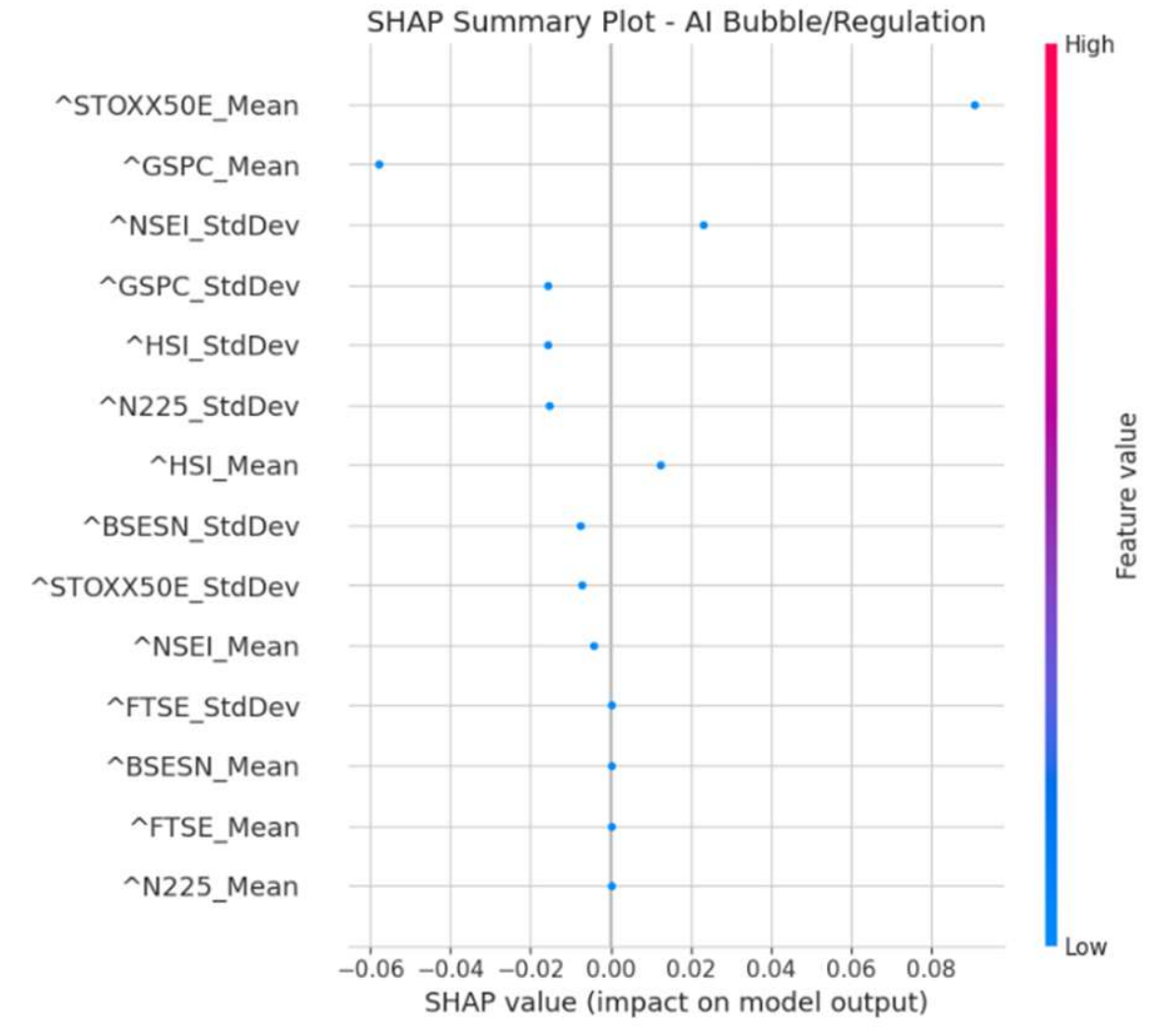

**Table 3: SHARC Decomposition Summary Across Scenarios**

| Scenario | Primary Loss Driver (Highest Negative SHAP) | Primary Mitigator (Highest Positive SHAP) | Mean vs. Variance Dominance | Terminal SVaR (%) |
|---|---|---|---|---|
| West Asia War | ^STOXX50E_Mean (Eurozone stagflation channel) | ^FTSE_Mean, ^N225_Mean (relatively decoupled) | Mean dominant during extreme shock | -2.2231 |
| Climate Risk | Geographically concentrated indices (HSI, STOXX50E) | U.S. S&P 500 (geographic diversification buffer) | Mean dominant under stress conditions; variance provides baseline | -2.1020 |
| AI Bubble / Regulation | ^GSPC_Mean (tech sector concentration) | Asian indices (lower AI sector weighting) | Mean dominant under stress conditions; variance provides baseline | -2.1599 |

## 4.3 Mean vs. Variance Dominance: The Core Finding

The primary analytical finding of this paper, and its most significant implication for capital limit-setting, is the non-linear risk decomposition pattern observed across all three scenarios. Under stress conditioning, the mean return component ($\varphi_{\mu_i}$) consistently dominates the variance component ($\varphi_{\sigma_i}$) in driving the final SVaR output.

### 4.3.1 Mechanism of Mean Dominance

Under normal market conditions, the GPR-predicted conditional variance ($\sigma_i$) is the primary input driving the VaR and ES calculation: a higher variance produces a wider predictive distribution and therefore a more negative quantile at the 99% confidence level. In this regime, variance dominates the decomposition, consistent with the theoretical perspective of McNeil et al. (2015) for the base case.

However, under stress conditioning, where the GPR has been fitted to a 252-day path of extreme negative returns, the relationship inverts. The directional magnitude of the stressed mean return ($\mu_i$), which becomes strongly negative under the imposed shock path, acts as the binding constraint on the capital measure. Intuitively, when a scenario dictates that an index must lose -40% over 252 days, the implied daily mean return is already deeply negative, and this directional loss outweighs the additional widening of the distribution driven by increased variance.

This mechanism is captured directly through the SHARC decomposition: $\varphi_{\mu_i}$ components are large and negative (pushing SVaR toward larger losses), while $\varphi_{\sigma_i}$ components remain present but smaller in magnitude. The relative contribution of mean and variance components varies by scenario but is consistently mean-dominated across the stress scenarios examined.

### 4.3.2 Regulatory Implications of Mean Dominance

The observed mean dominance under stress conditions has three direct regulatory implications.

First, it provides empirical grounding for the practitioner observation that 'volatility-based hedging may be ineffective during tail events': if the capital requirement is driven primarily by directional loss magnitude rather than volatility expansion, then variance-reducing hedges (such as options positions) may not reduce the SVaR-based capital requirement as effectively as directional position reductions.

Second, it validates the application of SHAP to SVaR rather than to base-case VaR. The mean dominance effect is specific to the stressed context, emerging because the scenario conditions the model on extreme directional losses. A SHAP analysis of base-case VaR would not reveal this pattern, as the base-case mean returns are approximately zero and the variance component would instead dominate the decomposition. The SVaR context is therefore not merely convenient, but the analytically appropriate setting for observing this behaviour.

Third, it provides a specific and actionable signal for capital limit management: when the SHARC decomposition indicates that the mean component contributes more than 80% of SVaR, risk managers should prioritise position reduction in the identified high-contribution indices rather than relying on volatility-based risk mitigation instruments.

## 4.4 The Force Plot as a Regulatory Audit Trail

The SHAP Force Plot provides the most directly actionable regulatory output of the analysis. For the West Asia War scenario with terminal SVaR of -2.2231%, the Force Plot decomposes this single capital number into its feature components, visually separating loss-amplifying and mitigating inputs, with magnitudes proportional to each feature's SHAP value.

Figure 4 presents the SHARC Force Plot for the West Asia War scenario and decomposition of terminal SVaR into baseline, mean-driven, and variance-driven components. Figures 5 and 6 present the corresponding force plots for the Climate Risk and AI Bubble scenarios, which exhibit similar decomposition patterns.

**Figure 4: SHARC Force Plot – West Asia War**

**Figure 5: SHARC Force Plot – Climate Risk**

**Figure 6: SHARC Force Plot – AI Bubble/Regulation**

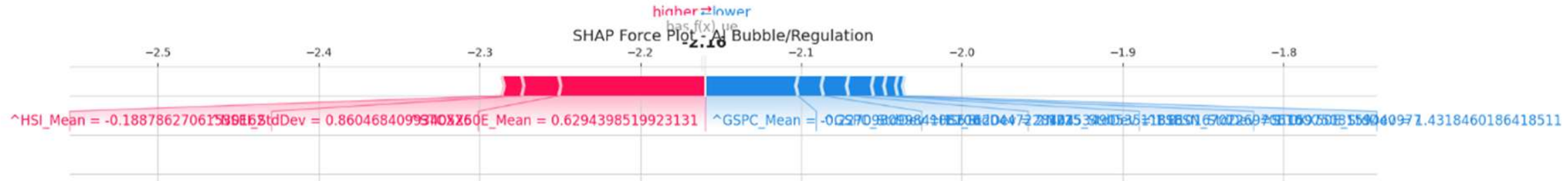

This visualization satisfies the FRTB requirement for auditable decomposition of capital outputs: a regulator examining this exhibit can immediately identify which index exposures are driving the capital requirement and by how much. The Force Plot is therefore not merely an analytical tool, but a regulatory documentation artifact that can be incorporated into ICAAP and CCAR model submissions.

Critically, the baseline component $\varphi_0$ represents the baseline SVaR under the SHARC decomposition, corresponding to the pre-stress capital level. The difference between this baseline and the terminal SVaR is then fully decomposed across the 14 input features, with zero remainder by the efficiency axiom. This completeness property ensures that no residual capital remains unexplained in the regulatory documentation.

# 5. Discussion

## 5.1 SHARC as a Production Risk Management Tool

The findings of this paper establish SHARC not merely as an academic explainability framework, but as a practical risk management tool with direct production applications. The translation from SHARC decomposition to management action requires operationalising the analytical results into decision rules that can be embedded within capital management workflows.

**Table 4: SHARC-Driven Risk Governance Applications**

| Application | SHARC Signal | Recommended Action |
|---|---|---|
| Capital Limit Setting | Mean component drives >80% of SVaR contribution | Prioritize position reduction over volatility-based hedging |
| Scenario Validation | SHAP ranking inconsistent with input shock magnitudes | Flag scenario design for review - potential transmission channel miscalibration |
| Regulatory Reporting | Force Plot decomposition of terminal day SVaR | Include as exhibit in ICAAP/CCAR model documentation |
| MLOps Monitoring | Variance component SHAP values shift significantly between reporting cycles | Trigger GPR kernel re-validation and hyperparameter review |
| Risk Governance | New scenario produces unexpected SHAP ordering | Escalate to model risk committee before capital submission |

The most operationally significant recommendation, using SHARC decomposition to guide capital limit-setting when the mean component contributes more than 80% of SVaR - represents a shift from static, volatility-anchored risk limits to dynamic, decomposition-informed limits. This shift is consistent with the direction of regulatory guidance: both ICAAP and CCAR frameworks increasingly emphasise the importance of understanding what drives capital requirements, not merely the level of capital itself.

## 5.2 Addressing the Scalability Challenge

A practical limitation of the local SHAP approach is the computational and cognitive overhead associated with generating and reviewing individual Force Plots for each capital calculation in a large, multi-asset portfolio. A production CCAR or ICAAP implementation may require SVaR calculations for hundreds of sub-portfolios, each requiring its own SHARC decomposition.

Three strategies address this scalability challenge. First, hierarchical SHAP application: apply Kernel SHAP at the portfolio level (as demonstrated in this paper) and use the portfolio-level decomposition to prioritise which sub-portfolio analyses require detailed review. Second, automated flagging: implement monitoring rules that trigger detailed SHAP analysis only when the mean-to-variance contribution ratio exceeds a predefined threshold, focusing human oversight on the highest-risk capital outputs. Third, SHAP aggregation: following Lundberg et al. (2020), aggregate local SHAP values into global importance summaries for routine reporting cycles, while reserving full local decomposition for regulatory submissions and model validation.

### 5.3 Relationship to Responsible AI Frameworks

The SHARC (SHAP for Regulatory Capital) framework documented in this paper is directly relevant to the emerging regulatory agenda around Responsible AI (RAI) in financial services. The EU AI Act (2024) mandates technical documentation including 'the logic involved' and 'the degree of autonomy' of the system. The SHARC framework provides the 'logic documentation' required: a complete, axiomatically grounded decomposition of each model output into its underlying input components.

Similarly, the Basel Committee's work on operational resilience (Basel Committee on Banking Supervision (BCBS), 2021) emphasises the need for financial institutions to understand and document the behaviour of automated systems used in risk management. The framework, through the use of SHAP Summary and Force Plots, provides the behavioural documentation required by these frameworks, establishing a bridge between technical ML explainability and operational risk governance.

### 5.4 The Local-Global Spectrum and Future Research

This paper deliberately occupies the 'local' end of the explainability spectrum, explaining specific capital outputs at specific points in time. The global end, which summarises average model behaviour across the full backtesting window, represents a complementary and equally important direction for model governance, particularly for internal model documentation and supervisory benchmarking.

Future research should extend the SHARC framework to global interpretation by aggregating local SVaR decompositions across all backtesting splits and all three scenarios to produce a comprehensive decomposition profile for the Hybrid GPR-HS framework. This would enable the framework's behaviour to be characterised not only for specific stress events but also as a general-purpose risk engine, supporting model approval processes that require evidence of consistent and predictable behaviour across a wide range of market conditions.

## 6. Conclusion

This paper resolves the primary regulatory barrier to adopting Gaussian Process Regression in ICAAP and CCAR capital workflows: the black box problem. By applying SHAP, a model-agnostic explainability method satisfying key axiomatic properties for fair and complete decomposition, to the SVaR outputs of the Hybrid GPR-HS framework, the paper demonstrates that the non-linear, non-parametric architecture of Vadrevu (2026a, 2026b) produces fully auditable and scenario-traceable capital estimates.

The three principal findings are as follows. First, the SHARC decomposition consistently maps GPR-HS SVaR outputs back to their scenario input design across all three macro scenarios, confirming framework fidelity: the model transmits scenario shocks faithfully through its non-linear architecture, and SHARC renders that transmission transparent and interpretable. Second, during tail events under stress conditioning, the mean return component (directional loss magnitude) dominates the variance component (volatility) in driving the final SVaR output, a non-linear decomposition pattern with direct implications for capital limit-setting and hedging strategy. Third, the SHAP Force Plot provides a regulatory documentation artifact, a complete, zero-remainder decomposition of any specific capital output, satisfying the audit trail requirements of FRTB, ICAAP Pillar 2, and CCAR model documentation standards.

Together with Vadrevu (2026a), which establishes the framework's predictive superiority and regulatory backtesting compliance and Vadrevu (2026b), which demonstrates its stability under forward-looking stress scenarios; this paper completes the regulatory architecture for the Hybrid GPR-HS framework: a model that is more accurate, more stable under stress, and more transparent than the parametric approaches it replaces.